# New strong sector, odd-parity processes, and the Tevatron

B. Holdom[1]

*Department of Physics*
*University of Toronto*
*Toronto, Ontario*
*Canada M5S 1A7*

Abstract

The color-octet isosinglet "rho" of a new strong- interaction sector is readily produced in $p\bar{p}$ collisions. Its odd-parity decay to an "eta" and a gluon may exceed its decay rate to dijets. At center of mass energies sufficiently greater than the colored "rho" mass, the odd-parity production of ("omega" + colored "eta") or ("rho" + colored "pion") may be comparable to $t\bar{t}$ production. Considering that the "omega" has a dominant odd-parity decay mode, we end up with ($Z$, $W$, or $\gamma$) + 4 jet events with two of the jets containing $b$ or $\bar{b}$ .

[1] holdom@utcc.utoronto.ca

In this paper we will consider the existence of a new strongly interacting sector characterized by a mass scale on the order of 200-400 GeV. New physics in this mass range is accessible to the Tevatron, and we would like to explore the relevant signatures. Such a sector of new physics may be associated with the generation of quark and lepton masses. The scale of this physics is somewhat below that required for electroweak symmetry breaking, but it is well known that a new strong gauge sector at 1 TeV (technicolor) is not by itself sufficient for generating quark and lepton masses. Additional dynamics is required, and it may exist on scales greater than 1 TeV (extended technicolor) and/or less than 1 TeV. Recent examples of the latter possibility are found in [1][2][3]. In particular, the phenomenological implications of "multiscale-technicolor" models have been studied in some detail [4].

The low-lying sector is distinguished from a conventional technicolor sector by its relative lack of participation in electroweak symmetry breaking. To emphasize this we will use the prefix "meta" rather than "techni".[2] The important point is that the strong production of metacolor resonances does not imply the strong production of the longitudinal components of the $W$ and the $Z$. In a multiscale technicolor model (here metacolor is technicolor) the light sector fields have small overlap with the Goldstone boson fields. Decays into longitudinal components of the $W$ and $Z$ are suppressed by powers of $F_1/F_2$, where $F_1$ and $F_2$ are the technipion decay constants of the light and heavy sector respectively [1]. In [3] the gauge interaction involved in electroweak symmetry breaking itself breaks at 1 TeV, and the unbroken subgroup is metacolor. Here the metacolor sector does not participate in electroweak symmetry breaking at all, and the Goldstone bosons due to electroweak breaking would couple to metafermions only through loops involving heavy fields. This would yield even weaker couplings of Goldstone bosons to metacolor resonances.

A requirement for our study is that the metafermions include at least one doublet which carries the same $SU(3)_C \times SU(2)_L \times U(1)$ quantum numbers as a doublet of quarks. We shall consider the mesonic resonances composed of these colored fermions, and we initially assume isospin symmetry in their description. The main difference between the spectrum of states in this sector and the QCD mass spectrum is that the ground-state pseudoscalar states may be relatively heavier (but still lighter than the vector mesons). $\rho_M$ and $P_M$ will denote the vector and pseudoscalar states, with a prime added for isosinglet states and a subscript "8" added for color octet states.

Also, we will be interested in processes in $p\bar{p}$ collisions with high partonic center-of-mass energy, such that at Tevatron energies the dominant production source will be $q\bar{q}$ annihilation and not $gg$ fusion. This is the situation for $t\bar{t}$ production.

Consider first the color-octet, isosinglet metarho, $\rho_{M8}^{0\,\prime}$. This is the resonance produced most copiously in $p\bar{p}$ collisions, since it is produced in $q\bar{q}$ annihilations via a virtual gluon. $\rho_{M8}^{0\prime}$ decays strongly to two colored metapions, if allowed, which in turn would most often produce 4 jets. If the $\rho_{M8}^{0\prime}$ mass is below twice the lightest colored metapion mass then the standard expectation is that its width is dominated by decay into two jets via $gg$ or $q\bar{q}$. But consider the "odd-parity" decay $\rho_{M8}^{0\prime} \to (P_M^{0\prime} g$ or $P_{M8}^{0\prime} g)$ where $P_M^{0\prime}$ ($P_{M8}^{0\prime}$) is the color-singlet (color-octet) metaeta. This width can be estimated by scaling up $\Gamma(\rho \to \pi^0\gamma)$, while the competing width to $gg$ or $q\bar{q}$ can be estimated by scaling up $\Gamma(\rho \to e^+e^-)$. We find, ignoring the phase space suppression factor,

$$\frac{\Gamma(\rho_{M8}^{0\prime} \to P_M^{0\prime} g \text{ or } P_{M8}^{0\prime} g)}{\Gamma(\rho_{M8}^{0\prime} \to q\bar{q} \text{ or } gg)} \approx \frac{1+5/2}{5/2+3/2} \frac{\alpha}{\alpha_s} \frac{9\Gamma(\rho \to \pi^0\gamma)}{\Gamma(\rho \to e^+e^-)} \approx 8.9 \ . \tag{1}$$

[2] We use the term metacolor in a more general sense than in [3].

The 5/2 in the numerator is the color factor for the $P_{M8}^{0\prime} g$ mode, the 5/2+3/2 includes flavor and color factors for the $q\bar{q}$ and $gg$ modes, the 9 compensates for the isosinglet photon coupling,[3] and we use $\alpha_s = 0.1$ for the value of the QCD coupling at the scale of the $\rho_{M8}^{0\prime}$ mass. The leading dependence on the unknown number of metacolors will cancel in the ratio.

Thus we find that the $\rho_{M8}^{0\prime}$ decay into $P_M^{0\prime} g$ or $P_{M8}^{0\prime} g$ can easily compete with the modes which produce dijets.[4] If $P_M^{0\prime}$ and $P_{M8}^{0\prime}$ are below the $t\bar{t}$ threshold then their dominant decays is to two gluons, and the result is a source of 3-jet events with total invariant mass peaked at the $\rho_{M8}^{0\prime}$ mass. Of course if the $P_M^{0\prime}$ and $P_{M8}^{0\prime}$ masses are too large then the resulting phase space suppression can imply that the 2-jet modes will dominate. The conclusion is that the production of the $\rho_{M8}^{0\prime}$ will yield 2, 3, and/or 4 jet events with a relative frequency determined by the mass spectrum. These are interesting signals, but they must compete against a large QCD background.

[The other case, when the $P_M^{0\prime}$ and/or $P_{M8}^{0\prime}$ is above the $t\bar{t}$ threshold, may be an interesting top quark production mechanism. This is similar to the $t\bar{t}$ production mechanism in [5] which also involves an intermediate $P_{M8}^{0\prime}$. One difference is that the mechanism in [5] relies on $gg$ fusion, while the mechanism here can proceed via $q\bar{q}$ annihilations. We will consider these issues in more detail elsewhere [6].]

Do signatures more striking than multijet production occur? One such signal is $Z$, $W$, or $\gamma$ production in association with four jets, with some jets containing a $b$ or $\bar{b}$. We will find new physics contributions to this signal when we consider the production of metacolor resonances for center of mass energies greater than the $\rho_{M8}^{0\prime}$ mass. To get some idea of what to expect we turn to the low energy data for $e^+e^-$ collisions, as shown in Fig. 1. (This figure is extracted from Ref. [7]). At $\sqrt{s} \approx 1.1$ GeV $\approx 1.4 m_\rho$, the production of 3 or more pions starts to exceed the production of 2 pions. At this $\sqrt{s}$ the ratio

$$R_1 \equiv \frac{\sigma[e^+e^- \to (\geq 3\,\text{pions})]}{\sigma[e^+e^- \to \mu^+\mu^-)]} \tag{2}$$

[3] We could have used $\Gamma(\omega \to \pi^0\gamma)$ in place of $9\Gamma(\rho \to \pi^0\gamma)$, but then there is a complication with $\omega$-$\phi$ mixing.

[4] We have reached a conclusion different from the one appearing in a footnote in [4].

is roughly 1/2.

Figure 1: $R \equiv \sigma[e^+e^- \rightarrow \text{hadrons}]/\sigma[e^+e^- \rightarrow \mu^+\mu^-]$

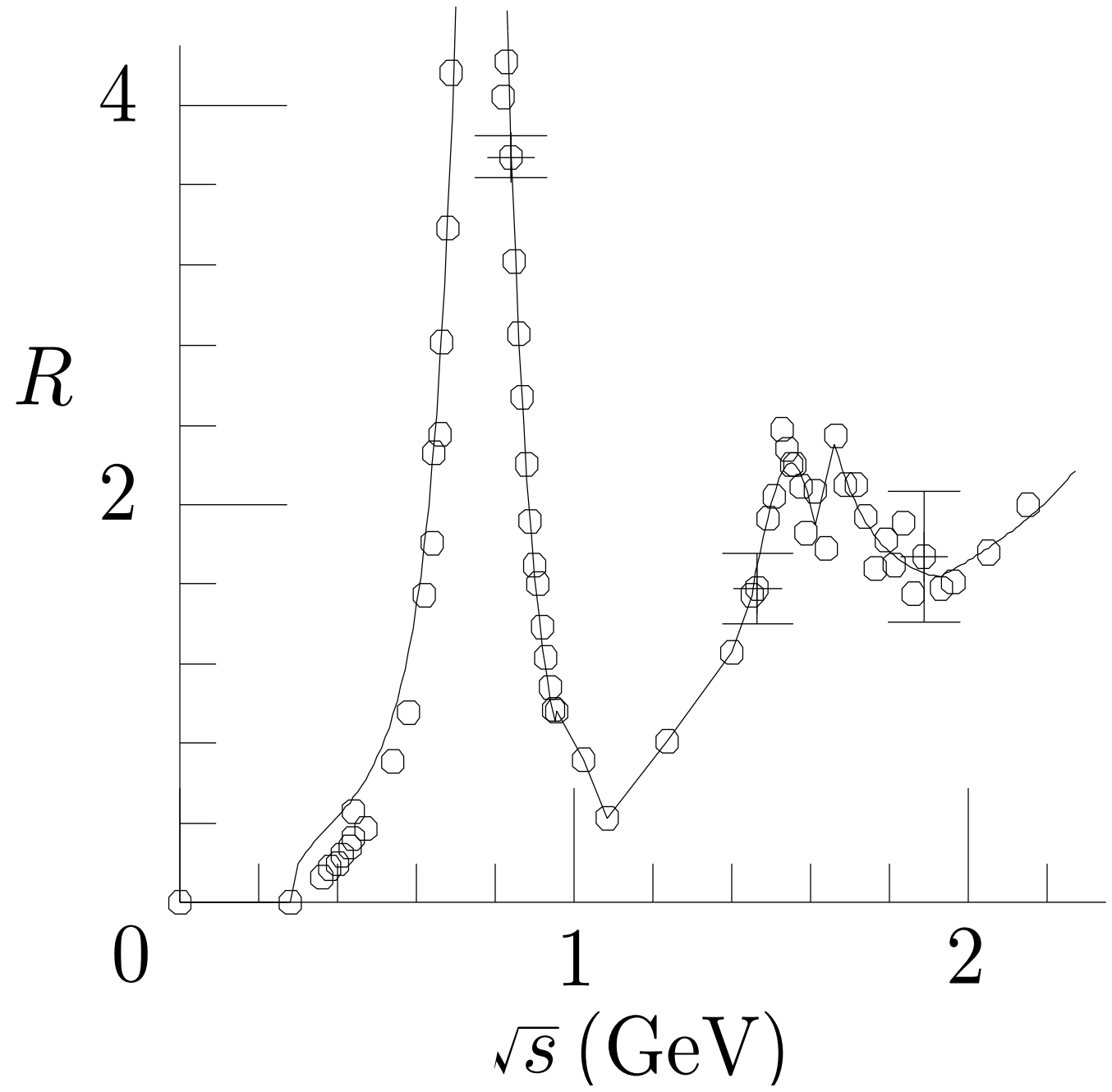

$R_1$ rises rapidly until it peaks at $\sqrt{s} \approx 1.6$ GeV, at which point it exceeds 2. The main source of data [8] between these two scales includes $4\pi$ but not $3\pi$ production, thus providing only a lower bound on $R_1$. In this region and even higher, the production of multipion states is dominated by quasi-two-body processes [9].

$$e^+e^- \rightarrow \gamma^\star \rightarrow m_1 + m_2 \rightarrow n\pi \tag{3}$$

The dominant contribution to the $4\pi$ data occurs via $\omega + \pi$, as evidenced by the rapid increase in $4\pi$ production occurring at the $\omega + \pi$ threshold [8][9]. The production of $\omega + \pi$ is well modeled by vector meson dominance via an intermediate off-shell $\rho^0$, which involves the odd-parity coupling $g_{\rho\omega\pi}$ [10].

In the metacolor case we may assume a similar occurrence; we may have the cross section for $q\bar{q} \rightarrow g^\star \rightarrow (\rho_M^{0\prime} P_{M8}^{0\prime}$ or $\rho_M^{\pm,0} P_{M8}^{\mp,0})$ starting to compete with the cross section for the production of a pair of colored metapions for sufficiently large partonic c.m. energy $\sqrt{\hat{s}}$. The required $\sqrt{\hat{s}}$ is probably higher than $1.4\, m(\rho_{M8}^{0\prime})$ since the pseudoscalars here are relatively heavier. $q\bar{q} \rightarrow \rho_M^{0\prime} P_{M8}^{0\prime}$ is the exact analog of $e^+e^- \rightarrow \omega\pi$ with isospin space exchanged for color space, and thus we consider

$$R_2 \equiv \frac{\sigma[q\bar{q} \rightarrow \rho_M^{0\prime} P_{M8}^{0\prime}]}{\sigma[q\bar{q} \rightarrow b\bar{b}]} \tag{4}$$

in lowest order QCD. The cross section for any one of the three other final states $\rho_M^{\pm,0} P_{M8}^{\mp,0}$

should be similar, and all these processes could again be modeled by vector meson dominance via an intermediate off-shell $\rho_{M8}^{0\prime}$. If we take $R_1$ defined in (2) to be dominated by the production of $\omega + \pi$, and if we take the $\sqrt{s}$ in $R_1$ and the $\sqrt{\hat{s}}$ in $R_2$ to be above the respective thresholds by the equivalent amounts, then we estimate

$$\frac{R_2}{R_1} \approx \frac{6}{5} N_M \ . \tag{5}$$

This accounts for various charge, color and flavor factors, and $N_M$ is the number of metacolors. There are of course other modes where the final meta-vector-meson is colored instead of the meta-pseudoscalar, and other modes where both metamesons are colored. If the meta-vector-meson is colored then its dominant decays are such that the result is a multijet event.

We now consider the ratio of physically relevant quantities

$$\frac{\sigma[p\bar{p} \ \rightarrow \ \rho_M^{0\prime} P_{M8}^{0\prime}]}{\sigma[p\bar{p} \ \rightarrow \ t\bar{t}]} \ , \tag{6}$$

where each cross section involves an integral over $\sqrt{\hat{s}}$. Remembering that $q\bar{q}$ is the dominant production source at Tevatron energies, we may thus deduce from (5) that this ratio is of order unity if the $\rho_M^{0\prime} + P_{M8}^{0\prime}$ threshold is in the 400 GeV range, *i.e.* not too much higher than the $t\bar{t}$ threshold.

Perhaps most interesting is that we have a mechanism for the production of the color-singlet, isosinglet metaomega $\rho_M^{0\prime}$. We will assume that the metaomega is below threshold for decay to three metapions. The odd-parity decays involving longitudinal $W$'s and $Z$'s, $\rho_M^{0\prime} \rightarrow (W_L^+ W_L^- Z_L \ , \ Z_L\gamma, \ Z_L Z \ , \ W_L W)$ were considered in [11] for conventional technicolor, but here these modes are suppressed. Instead, the odd-parity decay involving the color-singlet metapion $P_M$ rather than the Goldstone boson is likely the dominant mode. We therefore have the processes

$$\begin{array}{l} q\bar{q} \ \rightarrow \ g^\star \ \rightarrow \ \rho_M^{0\prime} \ + \ P_{M8}^{0\prime} \\ \qquad\qquad\qquad\quad \hookrightarrow \ P_M^0 \ + \ (Z \ \text{or} \ \gamma) \\ \qquad\qquad\qquad\quad \hookrightarrow \ P_M^\pm \ + \ W^\mp \end{array} \tag{7}$$

which involve the vector coupling of the (transverse) $W$ and $Z$.

In the first branch, the $P_{M8}^{0\prime}$ and $P_M^0$ decay predominantly to $gg$ and $b\bar{b}$ respectively.[5] This provides us with a source of ($Z$ or $\gamma$) + 4-jet events, with two of the jets originating in a $b$ or $\bar{b}$. The second branch produces $W$ + 4-jet events, but the nature of these events will depend on the mass of the $P_M^+$. If high enough it decays to $t\bar{b}$, and if not the $c\bar{s}$ mode should dominate. The ratio of the $\gamma : Z : W$ rates is, ignoring phase space suppression, $1 : (1 - 2s^2)^2/(2cs)^2 : 2/(2cs)^2 = 1 : 0.4 : 2.8$ where $s \equiv \sin\theta_W$. If $m(P_M)$ approaches $m(\rho_M^{0\prime}) - m_Z$ , the photon mode can come to dominate. We note that there is no resonance peak in the total invariant mass of these events, but there will be resonant peaks in various of the other invariant mass combinations.

[5] The $P_{M8}^{0\prime}$ may also have a nontrivial branching ratio to $b\bar{b}$ .

As noted above we also have the production of the color- singlet, isotriplet metarho $\rho_M^{0,\pm}$. The metarho will decay predominantly into $2P_M$ if allowed. More interesting [1] is when it is below this threshold, in which case it can decay into $P_M + W_L$ (or $Z_L$ in the case of $\rho_M^{\pm}$). But here again, decays involving longitudinal $W$ or $Z$ are suppressed in some model dependent manner. Two other processes are then worth checking. One is the decay into $P_M$ plus a transverse $W$ or $Z$. The transverse component can substitute for the longitudinal component because of the axial couplings of the $W$ and $Z$, and this will give a lower bound on the total production of $P_M + (Z$ or $W)$. We compare this to the $2P_M$ mode, ignoring for the moment the $P_M$ mass, and estimate

$$\frac{\Gamma(\rho_M \rightarrow P_M W_T)}{\Gamma(\rho_M \rightarrow P_M P_M)} \approx \frac{12\pi\alpha f_\pi^2}{s^2 m_\rho^2} \approx 0.017 \ . \tag{8}$$

Another mode for the $\rho_M^0$ is the odd-parity decay. But here we find

$$\frac{\Gamma(\rho_M^0 \rightarrow P_M^{0\prime}\gamma)}{\Gamma(\rho_M \rightarrow P_M P_M)} \approx \frac{9\Gamma(\rho \rightarrow \pi^0\gamma)}{\Gamma(\rho \rightarrow \pi\pi)} \approx 0.007 \ . \tag{9}$$

We therefore expect that regardless of the suppression of decay modes involving longitudinal $W$'s or $Z$'s, the dominant process involving the metarho $\rho_M^{0,\pm}$ is the following.

$$\begin{aligned} q\bar{q} \rightarrow g^\star \rightarrow\ & \rho_M + P_{M8} \\ & \hookrightarrow P_M + (W \text{ or } Z) \end{aligned} \tag{10}$$

In the $W$ mode one of $P_M$ or $P_{M8}$ is neutral and the other is charged (both are isotriplet). The nature of these events again depends on whether the charged metapion is above the $t\bar{b}$ threshold. In any case the $W$ + 4-jet events have at least two of the four jets as $b$-jets from the decay ($P_M^0$ or $P_{M8}^0$) $\rightarrow b\bar{b}$ . This produces a signal which is similar in most respects to the $W$ + 4-jet events with $b$-tags used as a signature of $t\bar{t}$ production at the Tevatron. In the $Z$ mode both $P_M$ and $P_{M8}$ are charged. If $P_{M8}^{\pm}$ is below the $t\bar{b}$ threshold then the result is $Z$ + 4-jets with none of the jets having $b$'s, unlike (7). If $P_{M8}^{\pm}$ is above the $t\bar{b}$ threshold then we have $Z + (\geq 4$ jets) with $b$-tags. The ratio of the $W^{\pm} : Z$ rates is $2 : 1$, and the total rate for the sum of modes in (10) is approximately 3 times that in (7).

Our analysis has relied on isospin symmetry in the new sector. An example of how this may occur concurrently with a large top mass is described in [3]. If there is substantial explicit isospin breaking then our results would change. In particular the $\rho_M^0$ and $\rho_M^{0\prime}$ states could mix such that the mass eigenstates become $\rho_{\bar{U}U}$ and $\rho_{\bar{D}D}$ .[4] Both of these states would have decays resembling those of the $\rho_M^0$ , and our source of $\gamma$ + 4-jet events with $b$-tags would be lost.

In summary we have described how a new strong sector below the electroweak symmetry breaking scale can give rise to ($W$, $Z$, or $\gamma$) plus 4-jet events with $b$-tags in $p\bar{p}$ collisions at a rate similar to $t\bar{t}$ production. Such events could occur with a total invariant mass in the 400-600 GeV range, of interest to the Tevatron. The various metamesons appearing in these processes are relatively narrow and should provide clear invariant mass resonant peaks. We encourage in particular a search for single photon plus multijet events to complement the study of $W$ or $Z$ plus multijet events.

## Acknowledgements

I thank M. Bando for a collaboration which led into this work, and I thank M.V. Ramana for his comments and his interest. I also thank P. Sinervo for correspondence and J. Terning and G. Triantaphyllou for discussions. This work was supported in part by the Natural Sciences and Engineering Research Council of Canada.